\documentstyle{article}
\begin{document}

\title{CLASSICAL LIMIT OF THE TRAJECTORY \\
REPRESENTATION OF QUANTUM MECHANICS,\\
LOSS OF INFORMATION\\
AND RESIDUAL INDETERMINACY}

\author{Edward R.\ Floyd \\
10 Jamaica Village Road, Coronado, CA 92118-3208, USA \\
floyd@crash.cts.com \\[.125in]
LANL E-Print Archive: {\bf quant-ph/9907092} [Ver. 3] \\[.125in] 
Accepted for publication in {\it Int.\ J.\ Mod. Phys.}\ {\bf A}}

\date{29 July 1999, revised 28 January 2000}

\maketitle

\begin{abstract}
The trajectory representation in the classical limit $(\hbar \to
0)$ manifests a residual indeterminacy.  We show that the
trajectory representation in the classical limit goes to neither
classical mechanics (Planck's correspondence principle) nor
statistical mechanics.  This residual indeterminacy is contrasted
to Heisenberg uncertainty.  We discuss the relationship between
residual indeterminacy and 't~Hooft's information loss and
equivalence classes.
\end{abstract}


\normalsize

\section{Introduction}  The transition from quantum motion to
classical motion has been a conceptually blurred area for the
various representations of quantum mechanics.  Recently, Faraggi
and Matone have shown that, while quantum mechanics is compliant
with the equivalence principle where all quantum systems can be
connected by an equivalence coordinate transformation (trivializing
map), classical mechanics is not.$^{\ref{bib:fm1}-\ref{bib:fm3}}$ 
Faraggi and Matone through the equivalence principle have
independently derived, free from axioms, the quantum  
Hamilton-Jacobi foundation for the trajectory representation of
quantum mechanics.  Herein, we theoretically test Planck's
correspondence principle$^{\ref{bib:planck}}$ by investigating the
transition from quantum to classical mechanics for the trajectory
representation, embedded in a non-relativistic quantum 
Hamilton-Jacobi theory, for the Planck constant going to zero in
the limit.

The quantum stationary Hamilton-Jacobi equation (QSHJE) is a
phenomenological equation just like the classical stationary
Hamilton-Jacobi equation (CSHJE) and the Schr\"{o}dinger equation. 
The QSHJE and CSHJE render generators of the motion that describe
trajectories that differ.  As we shall show herein, these
trajectories still differ in the classical limit.  The QSHJE in one
dimension is given for non-relativistic quantum motion
by$^{\ref{bib:fm1},\ref{bib:messiah},\ref{bib:prd34}}$ 

\begin{equation}
\frac{W_x^2}{2m}+V-E=-\frac{\hbar^2}{4m}\langle W;x\rangle
\label{eq:qshje}
\end{equation}

\noindent where $W$ is the reduced action (also known as Hamilton's
characteristic function), $W_x$ is the conjugate momentum, $V$ is
the potential, $\hbar=h/(2\pi )$, $h$ is the Planck constant, $m$
is the mass, and $\langle W;x\rangle$ is the Schwarzian derivative
that manifests the quantum effects.  The Schwarzian derivative
contains higher-order derivative terms given by

\[
\langle W;x\rangle = \frac{W_{xxx}}{W_x} - \frac{3}{2} \left(
\frac{W_{xx}}{W_x}\right)^2.
\]

\noindent  The complete solutions for the reduced action and
conjugate momentum are well known and given for energy $E$
by$^{\ref{bib:prd34},\ref{bib:afb20}}$

\begin{equation}
W = \hbar \arctan \left(\frac{b \theta /\phi + c/2}{(ab-
c^2/4)^{1/2}}\right) + K,\ \ \ \ x>0.
\label{eq:ra}
\end{equation} 

\noindent and

\begin{equation}
W_x=(2m)^{1/2}(a\phi ^2 + b\theta ^2 + c\phi \theta)^{-1}
\label{eq:cme}
\end{equation}

\noindent where $K$ is an integration constant that we may
arbitrarily set $K=0$ for the rest of this investigation and where
$(\phi ,\theta)$ is the set of independent solutions to the
associated stationary Schr\"{o}dinger equation for energy $E$.  The
Wronskian ${\cal W}(\phi ,\theta)$ is normalized so that ${\cal
W}^2 = 2m/[\hbar ^2(ab-c^2/4)]$.  The set of coefficients $(a,b,c)$
for energy $E$ determines the particular solution for $W$ and
$W_x$.  The set $(a,b,c)$ is determined by a sufficient set of a
combination of initial values or constants of the motion other than
$E$ for the third-order QSHJE.  This requires three independent
values.  For example, the set of coefficients $(a,b,c)$ can be
specified by the initial values $[W_x(x_o),W_{xx}(x_o)]$ for the
QSHJE plus the Wronskian, ${\cal W}$, which is a constant for the
Schr\"{o}dinger equation.  The particular solution was shown to
specify the particular microstate which had not been detected in
the Schr\"{o}dinger
representation.$^{\ref{bib:pla249},\ref{bib:prd34},
\ref{bib:carroll}}$  The left side of the QSHJE, Eq.\
(\ref{eq:qshje}), is the CSHJE.  If $\hbar $ were zero (as
distinguished from the limit $\hbar $ going to zero), then Eq.\
(\ref{eq:qshje}) would trivially reduce to the classical 
time-independent Hamilton-Jacobi equation.  Herein, we test
Planck's correspondence principle by investigating the proposition
that quantum mechanics transitions to classical mechanics in the
limit that $\hbar $ goes to zero.  We specify the limit $\hbar \to
0$ to be the classical limit.  (Another ``classical limit", neither
used nor investigated herein, is Bohr's correspondence principle
where the principle quantum number increases without
limit.$^{\ref{bib:bohr}}$)  The trajectory representation of
quantum mechanics is well suited for testing Planck's
correspondence principle.  While the CSHJE is a first-order
nonlinear differential equation, the QSHJE is a third-order
nonlinear differential equation whose second- and third-order terms
contain the factor $\hbar ^2$.  This investigation must include, in
the limit $\hbar \to 0$, the effects of annulling all higher than
first order terms in a differential equation upon the particular
solution and the set of necessary and sufficient initial values.  

With the trajectory representation, we show that quantum mechanics
in the classical limit does not generally reduce to classical
mechanics.  Nor does it reduce to statistical mechanics.  A
residual indeterminacy implies an uncertainty exists in general in
the classical limit.  We investigate this residual indeterminacy
and contrast it to Heisenberg uncertainty.  We also gain insight
into the quantum term $\hbar ^2\langle W;x \rangle /(4m)$.

Recently, 't~Hooft proposed that underlying contemporary quantum
mechanics there should exist a more fundamental, albeit still
unknown, theory at the Planck level that would provide more
information than the Schr\"{o}dinger wave
function.$^{\ref{bib:thooft}}$  In 't~Hooft's proposal, the
additional information distinguishes primordial states at the
Planck level, but this information is lost through dissipation as
these states evolve into states forming an equivalence class. 
't~Hooft suggested quantum gravity would dissipate information. 
States of an equivalence class, after a while, become
indistinguishable from each other even though they have different
pasts.  't~Hooft's ideas bear upon this investigation by giving us
insight into the relationship of the Copenhagen interpretation to
the trajectory representation where information regarding
microstates (primordial states) becomes ``lost" and where the
Schr\"{o}dinger wave function becomes an equivalence class.  We
compare residual indeterminacy with 't~Hooft's information loss and
equivalence classes.  This comparison is preliminary because the
underlying fundamental theory of 't~Hooft's proposal is still
incomplete.   

We examine the classical limit for three cases: a particle in the
classically allowed region; a particle in the classically forbidden
region (beyond the WKB turning point); and a particle in the
vicinity of the WKB turning point.  We choose potentials that are
heuristic and whose trajectory representation can be presented by
familiar functions.  One dimension suffices for this investigation.

As noted in the opening paragraph, Faraggi and
Matone,$^{\ref{bib:fm1}-\ref{bib:fm3}}$ in independently deriving
the QSHJE, have shown that such an equivalence principal does not
exist for classical mechanics (i.e. $\hbar =0$).  This begs the
question ``What happens when $\hbar \to 0$?"  Faraggi and Matone
essentially examined$^{\ref{bib:mpc}}$ the $\lim_{\hbar \to 0} W
\to W^{\mbox{\scriptsize classical}}$ in Ref.\ \ref{bib:fm3} (more
on the differences between their and our limiting processes in
Sec.\ 5).  Still, Faraggi and Matone found that the equivalence
principal in the classical limit is not valid.$^{\ref{bib:fm3}}$

Although the Bohmian school initially reported$^{\ref{bib:pr85}}$
that the classical limit was specified by $\lim_{\hbar \to 0}$, it
latter recanted and now opposes such a classification because
$\hbar$ is finite and a constant that cannot be
varied.$^{\ref{bib:pr144}}$  (The differences between the
trajectory representation and Bohmian mechanics has been discussed
elsewhere.$^{\ref{bib:fm1}-\ref{bib:fm3},\ref{bib:prd26}-
\ref{bib:pe7}}$)  In fact, $\hbar $ is finite but very small.  For
this exposition, we treat $\hbar $ not as a physical constant but
as an independent variable.  This allows us to study the
infinitesimal limit to show that even in the $\lim_{\hbar \to 0}$,
quantum trajectories do not in general reduce to classical
trajectories.

In Sec.\ 2, we investigate the classical limit of the trajectory
representation of quantum mechanics for a particle in the
classically allowed region.  In Sec.\ 3, we examine a trajectory in
the classical limit in the classical forbidden region.  We
investigate in Sec.\ 4 a trajectory as it transits between the
classically allowed and forbidden regions across the WKB turning
point.  In Sec.\ 5, we discuss the impact of the classical limit
upon the set of initial values necessary and sufficient to specify
the trajectory.  In Sec.\ 6, we present the relationship between
the this investigation and 't~Hooft's information loss and
equivalence classes.  In Sec.\ 7, we contrast residual
indeterminacy to Heisenberg uncertainty and distinguish the
classical limit of the trajectory representation from classical and
statistical mechanics. 

\section{The Classically Allowed Case}  We examine here a particle
in the classically allowed region, $E>V(x)$, in the limit that
$\hbar \to 0$.  In the limiting process, the quantum motion does
not approach the classical motion as $\hbar \to 0$ in spite of the
$\hbar ^2$ factor in the right side of the QSHJE.  Quantum physics
is more subtle.  

Let us consider a heuristic example.  We choose a free particle,
$V=0$, for finite $\hbar$.  An acceptable set of independent
solutions to the Schr\"{o}dinger equation for the free particle is
given by

\begin{equation}
\phi = [E(ab-c^2/4)]^{-1/4}\cos [(2mE)^{1/2}x/\hbar ] \ \ \ \
\mbox{and} \ \ \ \ \theta = [E(ab-c^2/4)]^{-1/4}\sin
[(2mE)^{1/2}x/\hbar].
\label{eq:isc}     
\end{equation}

\noindent The coefficients $(a,b,c)$ specify the particular
microstate for a specified energy $E$ in accordance with a
sufficient set of a combination of initial values and constants of
the motion other than energy.$^{\ref{bib:fpl9}}$ 

The reduced action, Eq.\ (\ref{eq:ra}), may be expressed for $V=0$
by

\begin{equation}
W = \hbar \arctan \left(\frac{b \tan [(2mE)^{1/2}x/\hbar ] +
c/2}{(ab-c^2/4)^{1/2}} \right).
\label{eq:raa}
\end{equation}

\noindent For $a=b$ and $c=0$, then $W = (2mE)^{1/2}x$ which
coincides with the classical reduced action.  For $a \ne b$ or $c
\ne 0$, l'H\^{o}pital's rule is used to evaluate $W$ in the
classical limit rendering

\begin{equation}
\lim_{\hbar \to 0} W =
\frac{2(ab-c^2/4)^{1/2}(2mE)^{1/2}x}{a+b+[(a-b)^2+c^2]^{1/2}\cos
\{[2(2mE)^{1/2}x/\hbar ]+\cot ^{-1}[c/(a-b)]\}}.
\label{eq:raal}
\end{equation}

\noindent The $1/\hbar $ factor in the argument of the cosine term
in the denominator on the right side of Eq.\ (\ref{eq:raa}) induces
in the classical limit an essential singularity in the cosine term. 
This essential singularity in turn induces an indetermination in
the reduced action, W, in the classical limit if $a \ne b$ or $c
\ne 0$.  The magnitude of this indeterminacy is a function of the
particular microstate as determined by the set of coefficients
$(a,b,c)$.  The phase shift of this indeterminacy is a function of
$c/(a-b)$.  This indefiniteness in $\lim_{\hbar \to 0} W$ does not
exist when $\hbar $ is finite.  This is consistent with Faraggi's
and Matone's findings that the equivalence principle was applicable
to quantum mechanics but not classical mechanics ($\hbar$ must not
be zero for equivalence).$^{\ref{bib:fm3}}$  We shall return latter
in this section after we have established the equations of motion
to show that this indefiniteness can be removed and that another
generator of the motion, Hamilton's principal function, $S$, for
quantum motion in the classical limit goes to the Hamilton's
principal function, $S^{\mbox{\scriptsize classical}}$, of
classical mechanics. 
 
The quantum conjugate momentum for $V=0$ can be expressed as

\begin{equation}
W_x=\frac{2(2mE)^{1/2}(ab-c^2/4)^{1/2}}{a+b+[(a-b)^2+c^2]^{1/2}\cos
\{[2(2mE)^{1/2}x/\hbar ]+\cot ^{-1}[c/(a-b)]\}}.
\label{eq:qcm}
\end{equation}

\noindent In the limit $\hbar \to 0$ the cosine term in the
denominator in Eq.\ (\ref{eq:qcm}) fluctuates with an
infinitesimally short wave length.  This induces in $W_x$ a
residual indeterminacy in the classical limit.  In the trajectory
theory, we know the microstates, but in classical mechanics, we do
not know the microstate because classical mechanics operates with
a reduced set of initial values insufficient to specify the
microstate. (The Copenhagen interpretation by precept forbids
knowledge of the microstate.)  Since $\hbar $ is very small but
finite, we must consider what happens in the limit that $\hbar \to
0$.  

In the classical limit, the cosine term in Eq.\ (\ref{eq:qcm})
becomes indefinite for $a\ne b$ or $c\ne 0$ even when we know the
microstate.  Our inability to evaluate $W_x$ in its classical limit
even with knowledge of the set of coefficients $(a,b,c)$ specifying
the particular microstate is a residual indeterminacy that is
hypothetical since in reality $\hbar$ is finite. For finite
$\hbar$, $W_x$ is always specified by $(E,a,b,c,x)$ in the
trajectory representation, but the Copenhagen interpretation denies
knowledge of $(a,b,c)$ while championing Heisenberg uncertainty.  

Nevertheless, we can evaluate its average momentum by averaging
$W_x$ over one cycle of the cosine term using standard integral
tables.$^{\ref{bib:d524}}$  We shall use this averaging process to
gain insight into the quantum term $\hbar ^2\langle W;x \rangle
/(4m)$.  Nothing herein implies that we are considering an ensemble
of identical microstates rather than a solitary microstate.  The
averaging process leads to

\begin{equation}
\Bigl\langle \lim_{\hbar \to 0} W_x\Bigr\rangle _{\mbox{\scriptsize
ave}} = \lim_{\hbar \to 0} \frac{(2mE)^{1/2}}{\hbar \pi }
\int_{\frac{-\hbar \pi }{(8mE)^{1/2}}}^{\frac{\hbar \pi
}{(8mE)^{1/2}}} W_x(E,a,b,c,x+x')\, dx'=
\frac{2(2mE)^{1/2}(ab-c^2/4)^{1/2}}{(a+b)[1-\frac{(a-b)^2-c^2}{(a
+b)^2}]^{1/2}} = (2mE)^{1/2}.
\label{eq:aqcm}
\end{equation}

\noindent [We have changed for this step the operational order of
evaluating the classical limit and averaging on the right side of
Eq.\ (\ref{eq:aqcm}) because the averaging domain is dependent upon
$\hbar$.  We continue this practice throughout.]  In the classical
limit, the average conjugate momentum is the classical momentum and
microstate information as specified by the set of coefficients
$(a,b,c)$ is lost.

The average for $\lim_{\hbar \to 0}W_x^2$ is given with the aid of
standard integral tables$^{\ref{bib:d534}}$ to be

\begin{equation}
\Bigl\langle\lim_{\hbar \to 0}W_x^2\Bigr\rangle_{\mbox{\scriptsize
ave}} = \Bigl\langle\Bigl(\lim_{\hbar \to
0}W_x\Bigr)^2\Bigr\rangle_{\mbox{\scriptsize ave}} = mE
\frac{a+b}{(ab-c^2/4)^{1/2}} \geq 2mE.
\label{eq:aqcm2}
\end{equation}  

\noindent  If we identify $W_x^2/(2m)$ as the effective kinetic
energy, then the average of the classical limit of the effective
kinetic energy for $V=0$ is greater than $E$ for $a \ne b$ or $c
\ne 0$ and is equal to $E$ for $a=b$ and $c=0$. 

Now let us examine the variance of $W_x$ in the classical limit. 
By Eqs.\ (\ref{eq:aqcm}) and (\ref{eq:aqcm2}), we have

\begin{equation}
\Bigl\langle \lim_{\hbar \to 0} W_x^2\Bigr\rangle
_{\mbox{\scriptsize ave}} - \Bigl( \Bigl\langle \lim_{\hbar \to 0} 
W_x\Bigr\rangle _{\mbox{\scriptsize ave}} \Bigr)^2 =
2mE\frac{a+b-(4ab-c^2)^{1/2}}{(4ab-c^2)^{1/2}} \geq 0.
\label{eq:vqcm}
\end{equation}

\noindent Even in the classical limit, the variance of the quantum
conjugate momentum, $W_x$, still is a function of the coefficients
$(a,b,c)$ that, in turn, manifest microstates.  For $a=b$ and
$c=0$, then the variance of $W_x$ is zero.  In this particular
microstate, the quantum motion reduces to classical motion for any
value of $\hbar$ because the additional necessary initial values of
the QSHJE, $[W_x(x_o),W_{xx}(x_o)]$ are both zero for a given
energy $E$.   
 
The average energy associated with the Schwarzian derivative term
of the QSHJE in the classical limit is given from Eqs.\
(\ref{eq:aqcm}--\ref{eq:vqcm}) by 

\begin{equation}
\Bigl\langle \lim_{\hbar \to 0} \frac{\hbar ^2}{4m} \langle
W;x\rangle \Bigr\rangle _{\mbox{\scriptsize ave}} =
E\left(1-\frac{(a+b)/2}{(ab-c^2/4)^{1/2}}\right) =
-\frac{\mbox{variance\ of\ }{\textstyle \lim_{\hbar \to 0}}W_x}{2m}
\le 0.
\label{eq:bqp}
\end{equation}

\noindent  So the average energy, in the classical limit, of the
quantum term, $\hbar ^2\langle W;x\rangle /(4m)$, which is also
known for unbound states as Bohm's quantum potential, $Q$, is
proportional to the negative of the variance of the classical limit
of the conjugate momentum.  The quantum potential is a function of
the particular microstate and may be finite even in the classical
limit as shown by Eq.\ (\ref{eq:bqp}).  As such, this potential is
not a function of spatial position alone but is path dependent and,
thus, cannot be a conservative potential.  Other objections to
Bohm's quantum potential have already been discussed
elsewhere.$^{\ref{bib:fm1}-\ref{bib:fm3},\ref{bib:prd26}-
\ref{bib:pe7}}$  Others have identified Q to be, even in a 
non-relativistic case, an internal motion, type of spin, or a
Zitterbewegung,$^{\ref{bib:pra57},\ref{bib:qp9902019}}$.  But
Carroll$^{\ref{bib:carroll}}$ has observed that this work assumes
a particle velocity inconsistent with Jacobi's theorem.  Let me
comment on this spin or Zitterbewegung.  If the quantum term is
associated with a spin or Zitterbewegung, this would imply that it
should manifest a kind of kinetic energy.  But such a kinetic
energy, even for a free particle, cannot be positive on average by
the Eq.\ (\ref{eq:bqp}).  This undermines the concept of kinetic
energy.  For particles in classically forbidden regions, such spin
or Zitterbewegung energy must be strongly negative.  This questions
attributing such motions to the Schwarzian derivative.

Let us now consider the equation of motion that is given by
Jacobi's theorem, $t-t_o=W_E$.  For the free particle with energy
$E$, the motion is given by

\begin{equation}
t-t_o=\frac{(ab-c^2/4)^{1/2}(2m/E)^{1/2}x}{a+b+(a^2-2ab+b^2+c^2)^
{1/2}\cos \{2(2mE)^{1/2}x/\hbar +\cot ^{-1}[c/(a-b)]\}}.
\label{eq:qeom}
\end{equation}

Let us now evaluate Hamilton's principal function, $S=W-Et$, in the
classical limit.  Since we have solved the equations of motion, we
are able to show that $S$ is the time integral of the Lagrangian of
classical mechanics, $L^{\mbox{\scriptsize classical}}$.  For
$V=0$, the classical Lagrangian is a constant give by
$L^{\mbox{\scriptsize classical}}=E$.  As the right sides of Eqs.\
(\ref{eq:raal}) and (\ref{eq:qeom}) are dynamically similar with
regard to $x$, the indeterminacy on the right side of Eq.\
(\ref{eq:raal}) can be removed and the reduced action in the
classical limit may be expressed as a function of time by
$\lim_{\hbar \to 0} W=2E(t-t_o)$.  Also, the classical limit of the
reduced action is independent of the set of coefficients $(a,b,c)$. 
Subsequently, Hamilton's principal function may be expressed by

\begin{equation}
\lim_{\hbar \to 0} S = E(t-t_o) = \int_{t_o}^t L^{\mbox{\scriptsize
classical}}\, dt = S^{\mbox{\scriptsize classical}}.
\label{eq:hpft}
\end{equation}

\noindent Note that Eq.\ (\ref{eq:hpft}) holds regardless of the
particular microstate.  The right side of Eqs.\ (\ref{eq:raal}) and
(\ref{eq:qeom}) also have similar form regarding the set of
coefficients $(a,b,c)$.  When we used the equation of motion, Eq.\
(\ref{eq:qeom}), to remove the indefiniteness on the right side of
Eq.\ (\ref{eq:raal}), we also annulled microstate specification.

Let us now return to the equation of motion.  As before with Eq.\
(\ref{eq:qcm}), in the limit $\hbar \to 0$ the cosine term in the
denominator above fluctuates with an infinitesimally short wave
length.  This induces a residual indeterminacy in $t(x)$.  Again as
before in the classical limit, we can evaluate the average (this
time for $t$ rather than for $W_x$) by averaging the right side of
the preceding equation over one cycle of the cosine term in the
denominator although here we have $x$ as a factor in the numerator. 
We again use standard tables$^{\ref{bib:d524}}$ where the $x$
factor in the numerator is fasted to a single value over the
infinitesimally short wavelength of the cosine term.  This leads to

\begin{equation}
\Bigl\langle \lim_{\hbar \to 0} (t-t_o)\Bigr\rangle
_{\mbox{\scriptsize ave}} =
\frac{[(ab-c^2/4)(2m/E)]^{1/2}x}{(a+b)[1-\frac{(a-b)^2-c^2}{(a+b)
^2}]^{1/2}} = \left(\frac{m}{2E}\right)^{1/2}x
\label{eq:aqeom}
\end{equation} 

\noindent independent of the coefficients $(a,b,c)$.  While
individual microstates have a residual indeterminacy in the
classical limit in its motion in the $[x,t]$ domain, this
indeterminacy is centered about the classical motion in the $[x,t]$
domain regardless of which particular microstate is specified. 
Nevertheless, the degree of indeterminacy is a function of the
microstate as specified by $(a,b,c)$.

\section{Classically Forbidden Case}  The QSHJE renders real
solutions, including conjugate momentum, for the trajectory in the
classically forbidden region.  On the other hand, the CSHJE has a
turning point at the WKB turning point.  If classical trajectories
were permitted beyond the turning point, the classical momentum
would become imaginary.  Let us now examine a particle in the
classically forbidden region, $E<V(x)$, in the classical limit,
$\hbar \to 0$.  Here, we choose

\[
V= \left\{ \begin{array}{lc}
               U>E, &  \ x\ge 0 \\
               0,      &  \ x<0
           \end{array}
     \right.
\]

\noindent representing an infinite step barrier.  In the
classically forbidden region, $x>0$, the set of independent
solutions to the associated Schr\"{o}dinger equation is given by

\[
\phi = \frac{\exp \{-[2m(U-E)]^{1/2}x/\hbar \}}{[(U-E)(ab-
c^2)]^{1/4}} \ \ \ \ \mbox{and} \ \ \ \ \theta = \frac{\exp
\{[2m(U-E)]^{1/2}x/\hbar \}}{[(U-E)(ab-c^2)]^{1/4}}.
\]  

\noindent From Eq.\ (\ref{eq:cme}), the conjugate momentum can be
expressed by 

\[
W_x = \frac{(ab-c^2/4)^{1/2}[8m(U-E)]^{1/2}}{a \exp \{[-2m(U-
E)]^{1/2}x/\hbar\} + b \exp \{[2m(U-E)]^{1/2}x/\hbar\} + c}.
\]

\noindent As $\hbar \to 0$, the $\exp \{[2m(U-E)]^{1/2}x/\hbar\}$
term in the denominator of the above increases without limit. 
Since its coefficient $b$ is finite real, we have in the forbidden
zone in the classical limit for the conjugate momentum that

\begin{equation}
\lim_{\hbar \to 0} W_x = 0
\label{eq:clcmb}
\end{equation}

\noindent regardless of either the particular microstate specified
by $(a,b,c)$ or where the particle is in the forbidden region as
long as it is in the forbidden region by a finite distance. 
However, the conjugate momentum is generally not the mechanical
momentum, i.e. $W_x \ne m \dot{x}.^{\ref{bib:fm3},\ref{bib:prd26}}$

The reduced action in the forbidden region is given by

\begin{equation}
W = \hbar \arctan \left(\frac{b \exp \{2[2m(U-E)]^{1/2}x/\hbar \}
+c/2}{(ab-c^2/4)^{1/2}} \right), \ \ \ \ x>0.
\label{eq:raf}
\end{equation} 
 
\noindent The classical limit for reduced action in the forbidden
region renders

\[
\lim_{\hbar \to 0} W \to h/4,\ \ \ \ x>0
\]  

\noindent independent of either the particular microstate or where
the particle is as long as $x$ is finite positive  As the reduced
action in the forbidden region is a constant, we arbitrarily choose
to use the principal value to evaluate $\arctan (\infty ) = \pi/2$
in the above equation.  This is consistent with the Maslov
index$^{\ref{bib:maslov}}$ which becomes exact for a piecewise
constant potential.

The equation of motion is given by Jacobi's theorem, $t-t_o=W_E$. 
For the particle with sub-barrier energy inside the step barrier,
the motion is given by 

\[
t-t_o = \frac{(ab-c^2/4)^{1/2}}{a \exp \{-2[2m(U-E)]^{1/2}x/\hbar
\} + b \exp \{2[2m(U-E)]^{1/2}x/\hbar \} +c}  \left(\frac{2m}{U-E}
\right)^{1/2}x, \ \ x>0.
\]

\noindent In the limit $\hbar \to 0$, the equation of motion for a
particle at a finite distance inside the step barrier is given by

\begin{equation}
\lim_{\hbar \to 0} (t-t_o) = 0, \ \ \mbox{for }x \mbox{ finite
positive}.
\label{eq:cleomb}
\end{equation}

\noindent Thus, the particle travels with infinite speed in the
classical limit.  This counterintuitive finding is consistent with
the findings regarding tunnelling that showed that dwell time
decreased with increasing $[2m(U-E)]^{1/2}/\hbar
$.$^{\ref{bib:afb20},\ref{bib:ap166},\ref{bib:qp9708007}}$  For
completeness, some insight has been already gained on this
counterintuitive phenomenon by examining a particle with oblique
incidence to the barrier.$^{\ref{bib:pe7},\ref{bib:qp9708007}}$ 
Such a particle was shown to have a trajectory that is not normal
to the iso-$W$ surface inside the barrier. In the classical limit,
it was shown that the particle's trajectory becomes imbedded in an
iso-$W$ surface.

The quantum term, $\hbar^2\langle W;x\rangle/(4m)$, on the right
side of Eq.\ (\ref{eq:qshje}) is given in the forbidden region
inside the step barrier by

\[
\begin{array}{rl}
\frac{\hbar^2}{4m}\langle W;x\rangle = (U-E) & \left(\frac{-2
\Bigl(a \exp \{-2[2m(U-E)]^{1/2}x/\hbar \} + b \exp \{2[2m(U-
E)]^{1/2}x/\hbar \}\Bigr)}{a \exp \{-2[2m(U-E)]^{1/2}x/\hbar \} +
b \exp \{2[2m(U-E)]^{1/2}x/\hbar \} + c}\right. \\[.25in]
     & - \left.\frac{\Bigl(a \exp \{-2[2m(U-E)]^{1/2}x/\hbar \} -
b \exp \{2[2m(U-E)]^{1/2}x/\hbar \}\Bigr)^2}{\Bigl(a \exp \{-
2[2m(U-E)]^{1/2}x/\hbar \} + b \exp \{2[2m(U-E)]^{1/2}x/\hbar \} +
c\Bigr)^2}\right), \ \ \ x>0.
   \end{array}
\]

\noindent In the classical limit, the quantum term becomes

\begin{equation}
\lim_{\hbar \to 0} \left(\frac{\hbar^2}{4m}\langle W;x\rangle
\right) = E-U \leq 0, \ \ \ \ x>0.
\label{eq:clqpb}
\end{equation}

\noindent Equations (\ref{eq:clcmb}) and (\ref{eq:clqpb}) balance
the QSHJE, Eq.\ (\ref{eq:qshje}) in the classically forbidden
region in the classical limit as expected.

\section{WKB Turning Point}  We now investigate the trajectory in
the classical limit in the vicinity of the WKB turning point.  We
also examine the transition between the classically allowed and
classically forbidden regions in the classical limit as the
trajectory transits the WKB turning point.  We choose the potential
to be 

\begin{equation}
V=fx,
\label{eq:vtp}
\end{equation}

\noindent which represents a constant force $f>0$ acting on our
particle.  Any well-behaved one-dimensional potentials for which
the force remains finite and continuous can always be approximated
by a linear potential in a sufficiently small region containing the
WKB turning point.

Let us digress briefly.  In the previous two sections, we examined
potentials that were at least piecewise constant.  Even though the
independent solution set, $(\phi ,\theta )$, for a step potential
is mathematical simpler, such a potential does not have a classical
short-wave correspondence at the turning point for the relative
change in the potential over a wavelength remains large
there.$^{\ref{bib:ajp60}}$  The reflection coefficient is
independent of $\hbar $.  Even a sixteen-inch armor-piercing naval
projectile with super barrier energy would still experience partial
backscatter.

For a particle with energy $E$, the WKB turning point, $x_t$, is
given by $x_t=E/f$.  An acceptable set of independent solutions,
$(\phi ,\theta )$, to the Schr\"{o}dinger equation is formed  
of Airy functions given by

\begin{equation}
\phi = \frac{(2m)^{1/12} \pi^{1/2} \mbox{Ai}[(2mf/\hbar
^2)^{1/3}(x-E/f)]}{(\hbar f)^{1/6}(ab-c^2/4)^{1/2}} \ \ \ \
\mbox{and} \ \ \ \ \theta = \frac{(2m)^{1/12} \pi^{1/2}
\mbox{Bi}[(2mf/\hbar ^2)^{1/3}(x-E/f)]}{(\hbar f)^{1/6}(ab-
c^2/4)^{1/2}}.
\label{eq:isl}
\end{equation}

The reduced action for the linear potential specified by Eq.\
(\ref{eq:vtp}) is given by

\begin{equation}
W = \hbar \arctan \left(\frac{b \mbox{Bi}(\xi /\hbar
^{2/3})/\mbox{Ai}(\xi /\hbar ^{2/3}) + c/2}{(ab-c^2)^{1/2}}\right)
\label{eq:qhcft}
\end{equation}

\noindent where $\xi = (2mf)^{1/3}(x-E/f)$ for expository
convenience.  For $(fx-E)$ finite negative, the classical limit of
the reduced action in the classically allowed region outside any
infinitesimal neighborhood containing the WKB turning point is
given by

\begin{equation}
\lim_{\hbar \to 0} W=\frac{(4/3)(ab-c^2)^{1/2}(2m)^{1/2}}{a+b +[(a-
b)^2+c^2]^{1/2}\sin [\frac{4}{3} \xi ^{3/2}/\hbar+\cot^{-
1}(\frac{c}{a-b})]} \frac{(E-fx)^{3/2}}{f}, \ \ \ \ \mbox{for $(fx-
E)$ finite negative.}
\label{eq:qhcfta}
\end{equation} 

\noindent If $a=b$ and $c=0$, then Eq.\ (\ref{eq:qhcfta})
simplifies to

\[
\lim_{\hbar \to 0}W\bigg|_{a=b,c=0}=\frac{[2m(E-fx)]^{3/2}}{3mf},
\ \ \ \ \mbox{for $(fx-E)$ finite negative},
\]

\noindent which is consistent with the reduced action for classical
mechanics for the corresponding linear potential as expected.  For
$(fx-E)$ finite positive and for any microstate, then the classical
limit of the reduced action in the classically forbidden region
outside any infinitesimal neighborhood containing the WKB turning
point is constant given by

\[
\lim_{\hbar \to 0}W=h/4, \ \ \ \ \mbox{for $(fx-E)$ finite
positive},
\]

\noindent which is a constant consistent with the Maslov
index.$^{\ref{bib:maslov}}$

The conjugate momentum for the linear potential specified by Eq.\
(\ref{eq:vtp}) is given by

\[
W_x=\frac{(2\hbar mf)^{1/3}(ab-c^2/4)^{1/2}}{\pi [a \mbox{Ai}^2(\xi
/\hbar ^{2/3}) + b \mbox{Bi}^2(\xi /\hbar ^{2/3}) + c \mbox{Ai}(\xi
/\hbar ^{2/3}) \mbox{Bi}(\xi /\hbar ^{2/3})]}.
\]

\noindent For $(fx-E)$ finite negative, the classical limit of the
conjugate momentum in the classically allowed region outside any
infinitesimal neighborhood containing the WKB turning point is
given by 

\begin{equation}
\lim_{\hbar \to 0}W_x= \frac{2(ab-c^2/4)^{1/2} [2m(E-
fx)]^{1/2}}{a+b +[(a-b)^2+c^2]^{1/2}\sin [\frac{4}{3} \xi
^{3/2}/\hbar +\cot^{-1}(\frac{c}{a-b})]}, \ \ \ \mbox{for $(fx-E)$
finite negative}.
\label{eq:cmacl}
\end{equation}

\noindent For $a=b$ and $c=0$, then $\lim_{\hbar \to 0}W_x=[2m(E-
fx)]^{1/2}$, which is consistent with the classical momentum in the
allowed region.  For $(fx-E)$ finite positive and for any
microstate, the classical limit of the reduced action in the
classically forbidden region outside any infinitesimal neighborhood
containing the WKB turning point is zero as expected.

We now consider the quantum equation of motion that is given by
Jacobi's theorem.  For the particle with energy $E$ and subject to
the linear potential, Eq.\ (\ref{eq:vtp}), the motion is given by

\begin{equation}
t-t_o=\frac{\hbar ^{1/3}}{\pi }\frac{(ab-
c^2/4)^{1/2}(2m/f^2)^{1/3}}{a \mbox{Ai}^2(\xi /\hbar ^{2/3}) + b
\mbox{Bi}^2(\xi /\hbar ^{2/3}) + c \mbox{Ai}(\xi /\hbar ^{2/3})
\mbox{Bi}(\xi /\hbar ^{2/3})}.
\label{eq:qeomtp}
\end{equation}

\noindent For $(fx-E)$ finite negative, the classical limit of the
quantum equation of motion in the classically allowed region
outside any infinitesimal neighborhood containing the WKB turning
point is given by 

\begin{equation}
\lim_{\hbar \to 0}(t-t_o) = \frac{2(ab-c^2/4)^{1/2}[2m(E-
fx)]^{1/2}/f}{a+b +[(a-b)^2+c^2]^{1/2}\sin [\frac{4}{3} \xi
^{3/2}/\hbar +\cot^{-1}(\frac{c}{a-b})]}, \ \ \ \ \mbox{for $(fx-
E)$ finite negative}.
\label{eq:qeomtpa}
\end{equation}

\noindent  For the microstate specified by $a=b$ and $c=0$, the
equation of motion simplifies to

\[ 
\lim_{\hbar \to 0}(t-t_o)\bigg|_{a=b;c=0} = \frac{[2m(E-
fx)]^{1/2}}{f}, \ \ \ \ \mbox{for $(fx-E)$ finite negative},
\]

\noindent which is consistent with the classical equation of motion
for a linear potential.  In the forbidden region outside any
infinitesimal region containing the WKB turning point, the equation
of motion renders

\[
\lim_{\hbar \to 0}(t-t_o) = 0, \ \ \ \  \mbox{for $(fx-E)$ finite
positive},
\]

\noindent which is consistent with the classical forbidden region.

In summary, we see that in the classical limit there is a
consistent transition across a WKB turning point between the
classically allowed and forbidden regions.  Nevertheless, the
motion remains ``quantum" in an infinitesimal region containing the
WKB turning point.  However, the measure of this region in the
classical limit reduces to zero.

\section{Initial Values}  The QSHJE is a third order nonlinear
differential equation while the CSHJE is first order.  The reduced
action, $W$ or $W^{\mbox{\scriptsize classical}}$, does not
explicitly appear in either the quantum or classical equation
respectively.  Then, a set of necessary and sufficient initial
values at $x_o$ needed to specify the quantum conjugate momentum in
addition to the constant of motion $E$ is $[W_x(x_o),W_{xx}(x_o)]$
while just $E$ is necessary and sufficient to specify the classical
conjugate momentum.  Knowing $W(x_o)$ specifies the integration
constant $K$ in Eq.\ (\ref{eq:ra}).  Subsequently, the quantum and
classical reduced actions are known to within an arbitrary
integration constant.

The initial values for the solution of the QSHJE specify the
particular microstate.$^{\ref{bib:prd34},\ref{bib:fpl9}}$  We note
that, when $a=b$ and $c=0$, the quantum trajectory solution for
$V=0$ and for $V=fx$ in the classical limit reduced to the
classical trajectory solution.  The underlying physics is that the
coefficients $a=b$ and $c=0$ for $V=0$ render by Eq.\
(\ref{eq:qcm}) that $W_x$ be a constant, $(2mE)^{1/2}$, independent
of $\hbar $ consistent with the corresponding
$W_x^{\mbox{\scriptsize classical}}$.  Even for finite $\hbar $,
selecting $a=b$ and $c=0$ causes $W_{xx}$ and $W_{xxx}$ to be zero
consistent with classical mechanics for $V=0$.  If $E$ is unknown,
then $[W_x(x_o),W_{xx}(x_o),W_{xxx}(x_o)]$ forms a necessary and
sufficient set of initial values to specify
$W(x)$.$^{\ref{bib:prd29}}$  For the linear potential $V=fx$,
selecting $a=b$ and $c=0$, we have by Eq.\ (\ref{eq:cmacl}) and its
differentiation that $W_x=[2m(E-fx)]^{1/2}$ and  $W_{xx}=f$ again
consistent with classical mechanics for a linear potential.  Thus,
choosing $a=b$ and $c=0$ tacitly induces the necessary initial
values $[W_x(x_o),W_{xx}(x_o)]$ for the QSHJE to correspond to the
superfluous initial values $[W_x^{\mbox{\scriptsize
classical}}(x_o),W_{xx}^{\mbox{\scriptsize classical}}(x_o)]$ for
the corresponding CSHJE.  Another view is that classical mechanics
inherently assumes that $a=b$ and $c=0$ for our choices, Eqs.
(\ref{eq:isc}) and (\ref{eq:isl}), for the set $(\phi ,\theta )$ of
independent solutions.  For completeness, had we chosen a different
set of independent solutions than those specified by Eqs.\
(\ref{eq:isc}) or (\ref{eq:isl}) for $V=0$ and $V=fx$ respectively,
then, for that potential, we would have had to choose a different
set of coefficients $(a,b,c)$ to achieve Planck correspondence to
classical mechanics.

Let us now consider a special class of microstates.  For this
class, we set $(a-b)^2+c^2=\eta (\hbar ,\chi )$ for the sets of
independent solutions given by Eqs.\ (\ref{eq:isc}) and
(\ref{eq:isl}).  While $\hbar $ is treated as an independent
variable in this exposition, $\chi $ is not but rather is a set of
other physical constants.  We use this notation to make it explicit
that $\eta $, to be dimensionless, must be dependent upon other
physical constants.  Mathematically, $\eta $ is the square of the
amplitude of the trigonometric term, whose argument has inverse
$\hbar $ dependence, in the solutions for $W_x$ as shown by Eqs.\
(\ref{eq:qcm}) and (\ref{eq:cmacl}).  Also let us set $\lim_{\hbar
\to 0}\eta = 0$.  Then the classical limit of the trajectory
representation for this special class of microstates would have
Planck correspondence$^{\ref{bib:planck}}$ to classical mechanics
including correspondence to superfluous initial values
$[W_x(x_o),W_{xx}(x_o)]$.  For this class of microstates, the
limiting procedure with respect to $\hbar $ would be
analogous$^{\ref{bib:mpc}}$ to the procedure used by Faraggi and
Matone in Ref.\ \ref{bib:fm3}.

\section{Loss of Information}  Passing to the classical limit
incurs a loss of information, associated with a set of necessary
and sufficient initial values, due to going from the third-order
QSHJE to the first-order CSHJE.  Likewise, even for finite $\hbar
$, the Copenhagen interpretation looses the information inherent to
microstates of the trajectory representation despite the Copenhagen
school asserting that the Schr\"{o}dinger wave function is
exhaustive.  For completeness, we study loss of information for
both cases and compare our results with
't~Hooft.$^{\ref{bib:thooft}}$  There are similarities and
significant differences.

First, we shall examine the quantum level ($\hbar $ finite).  As
already noted, the trajectory representation in its Hamilton-Jacobi
formulation manifests microstates not discernible by the
Schr\"{o}dinger representation showing that the Schr\"{o}dinger
wave function cannot be the exhaustive description of
nature.$^{\ref{bib:pla249},\ref{bib:prd26},\ref{bib:fpl9}}$  For a
specified energy $E$, each microstate, by itself, specifies the
Schr\"{o}dinger wave function.  Yet, each microstate of energy
eigenvalue $E$ has a distinct trajectory specified by the set of
initial values $[W_x(x_o),W_{xx}(x_o)]$.  Are these microstates
primordial at the Planck level?  Yes, we have already shown
elsewhere$^{\ref{bib:prd29}}$ that the trajectories are
deterministic and that the trajectory representation has arbitrary
initial conditions without concern for any nonequilibrium initial
values that encumbered Bohm$^{\ref{bib:pr85},\ref{bib:pr144}}$ at
the Planck level.  As the Copenhagen school asserts that $\psi $
should be the exhaustive description of natural phenomenon, the
Copenhagen school denies that primordial microstates could exist. 
Viewed externally, the Copenhagen school unwittingly makes $\psi $
to be a de facto equivalence class of any putative microstates. 
This is consistent with the QSHJE being more fundamental than the
stationary Schr\"{o}dinger equation, in contrast to Messiah's
assertion,$^{\ref{bib:messiah}}$ because, as shown
elsewhere,$^{\ref{bib:fpl9}}$ the bound-state boundary conditions
of the QSHJE do not generate a unique solution but rather generate
an infinite number of ``primordial" microstates while the boundary
conditions for the Schr\"{o}dinger equation do generate a unique
``equivalence-class" bound-state wave function.  Let us make a few
comparisons with a 't~Hooft process.  The primordial microstates
are deterministic trajectories of discrete energy $E$ in contrast
to the 't~Hooft primordial states that are of the continuum. 
Nevertheless, as all initial values are allowed for the
microstates,$^{\ref{bib:prd29}}$ the trajectories for any
equivalence class manifested by $\psi $ densely spans finite phase
space.  Also, the Copenhagen school looses information on
primordial microstates by default and not through a 't~Hooft
dissipative process.      

Second, let us examine loss of information in the trajectory
representation by executing the classical limit. In the trajectory
representation, residual indeterminacy manifests loss of
information.  The residual indeterminacy for the solution, $W_x$,
of the QSHJE in the classical limit is given for $V=0$ and $V=fx$
by the trigonometric terms

\[
[(a-b)^2+c^2]^{1/2}\cos \{[2(2mE)^{1/2}x/\hbar]+\cot ^{-1}[c/(a-
b)]\}
\]

\noindent and

\[
[(a-b)^2+c^2]^{1/2}\sin \{\frac{4}{3}\xi ^{3/2}/\hbar +\cot ^{-
1}[c/(a-b)]\}
\]

\noindent in the denominators of Eqs.\ (\ref{eq:qcm}) and
(\ref{eq:cmacl}) respectively.  Here, we set the square of the
amplitude of these trigonometric terms to be $A = (a-b)^2+c^2$
where A is dimensionless.  Explicitly, A is specified to be
independent of $\hbar $ in contrast to $\eta $ of Sec.\ 5.  The
phase shift of the argument of these trigonometric terms is
manifested by $\cot ^{-1}[c/(a-b)]$.  The factor $\hbar ^{-1}$ in 
the argument of these trigonometric terms induces an indeterminacy
in the classical limit that makes the phase shift due to $\cot ^{-
1}[c/(a-b)]$ irrelevant.  This represents a loss of information. 
On the other hand, The phase shift, $\cot ^{-1}[c/(a-b)]$, is not
redundant for specifying the set coefficients $(a,b,c)$ from the
set of necessary and sufficient initial values
$[W_x(x_o),W_{xx}(x_o)]$ for the QSHJE for a given $E$.  Hence, the
irrelevance of the phase shift gives the set of coefficients
$(a,b,c)$ another degree of freedom that makes the set of
coefficients underspecified in the classical limit.  This
underspecification of coefficients $(a,b,c)$ permits the primordial
microstates to form into equivalence classes where the primordial
microstates establish the membership within any particular
equivalence class and become identical with one another in the
classical limit.  This information loss differs with that of
't~Hooft.  As before, the primordial microstates have discrete
rather than continuum energies.  Also again, no dissipation of
information occurs in the trajectory representation when going to
the classical limit, but rather this loss of information induces an
indeterminacy.  

We may generalize to say that as classical mechanics has a smaller
set of necessary and sufficient initial values than the trajectory
representation of quantum mechanics, then there is some loss of
information and the formation of equivalence classes as we go to
the classical limit.  Also, the Copenhagen school by precept
considers $\psi $ to be exhaustive and disregards any microstate
information.  In either case, this loss of information is not due
to any dissipation as it is in 't~Hooft's proposal.  Without
dissipation, this loss of information may occur for stationarity of
the quantum Hamilton-Jacobi equation.

\section{Indererminacy}  Let us begin by contrasting the residual
indeterminacy of the trajectory representation to Heisenberg
uncertainty.  Residual indeterminacy and Heisenberg uncertainty
manifest themselves in different regimes of $\hbar $.  In the
classical limit, $\hbar \to 0$, Heisenberg uncertainty goes to zero
as the commutation relations are linear in $\hbar $.  On the other
hand, the residual uncertainty in the trajectory representation
exists only for $\hbar \to 0$.  We note that residual indeterminacy
is consistent with the findings of Faraggi and Matone that the
equivalence principle exists for quantum mechanics but not for
classical mechanics.$^{\ref{bib:fm3}}$  Otherwise, the trajectory
representation remains causal$^{\ref{bib:ijmpa14}}$ and
deterministic.  

Heisenberg uncertainty exists in the $[x,p]$ domain (where $p$ is
momentum) since the Hamiltonian operates in the $[x,p]$ domain.  On
the other hand, the trajectory representation through a canonical
transformation to its Hamilton-Jacobi formulation operates in the
$[x,t]$ domain.$^{\ref{bib:park}}$  Residual indeterminacy of the
trajectory representation is in the $[x,t]$ domain, cf. Eqs.\
(\ref{eq:aqeom}) and (\ref{eq:qeomtpa}).

Heisenberg uncertainty, which denies the existence of simultaneous
knowledge of position and momentum, tacitly implies that the
existence of such knowledge would suffice to render determinism
even without knowing energy.  Bohm, in his salad days as a
Copenhagenist, cited the Heisenberg uncertainty principle for
denying determinism.$^{\ref{bib:bohm-qm}}$  Bohm had also noted
that knowledge of the exact position and velocity of an electron at
some particular time and knowledge of the forces acting on the
electron at all times would render the electron's classical
trajectory and classical determinism.$^{\ref{bib:bohm-qm-2}}$ 
While the set $[p(t_o),x(t_o)]$ is sufficient to render $E$ and
subsequently a unique solution (i.e., determinism) in classical
mechanics, it is an insufficient set of initial values for
specifying a unique solution for the QSHJE.  The QSHJE, as a 
third-order differential equation, requires knowing the initial
values of the higher order derivatives to specify a unique
solution.$^{\ref{bib:prd29}}$  As already noted herein, if $E$ is
unknown, then the set of initial conditions
$[W_x(x_o),W_{xx}(x_o),W_{xxx}(x_o)]$ for finite $x_o$ is necessary
and sufficient to determine $E$ and a unique solution, $W_x(x)$,
for the QSHJE.$^{\ref{bib:prd29}}$  The Heisenberg uncertainty
principle masks the fundamental cause of indeterminism in the
Copenhagen interpretation.  As long as the Copenhagen
interpretation, even without Heisenberg uncertainty, assumes an
insufficient set of initial values, tacitly or otherwise, to tackle
the QSHJE or any other representation capable of rendering quantum
motion, it has forfeited any chance to determine a unique solution. 
Without a sufficient set of initial values, it is premature to
postulate any uncertainty that may exist within the insufficient
set $[p(t_o),x(t_o)]$.

In closing, we remark on the impact of residual indeterminacy.  In
the classical limit, $\hbar \to 0$, the trajectory representation
of quantum mechanics does not generally go to classical mechanics
invalidating Planck's correspondence principle.  Nor does it go to
statistical mechanics as the amplitude of the indeterminacy is
given by $[(a-b)^2+c^2]^{1/2}$ for the sets of independent
solutions of the Schr\"{o}dinger equation used herein, cf. Eqs.\
(\ref{eq:raal}), (\ref{eq:qhcfta}), (\ref{eq:cmacl}), and
(\ref{eq:qeomtpa}).
 
\begin{center}
{\bf Acknowledgement}
\end{center}

I heartily thank M. Matone, A. E. Faraggi, and R. Carroll for their
stimulating discussions and encouragement.  I thank G. 't~Hooft for
discussing his ideas on information loss and equivalence classes. 
I also thank D. M. Appleby for his critique of an early manuscript.

\paragraph{References}
\begin{enumerate}\itemsep -.06in
\item \label{bib:fm1} A.\ E.\ Faraggi and M.\ Matone, {\it Phys.\
Lett.}\ {\bf B 437}, 369 (1998), hep-th/9711028.
\item \label{bib:pla249} A.\ E.\ Faraggi and M.\ Matone, {\it
Phys.\ Lett.}\ {\bf A 249}, 180 (1998), hep-th/9801033.
\item \label{bib:plb445} A.\ E.\ Faraggi and M.\ Matone, {\it
Phys.\ Lett.}\ {\bf B 445}, 357 (1998), hep-th/9809126; {\it Phys.\
Lett.}\ {\bf B 450}, 34 (1999), hep-th/9705108.
\item \label{bib:fm3} A.\ E.\ Faraggi and M.\ Matone, {\it Phys.\
Lett.}\ {\bf B 445}, 77 (1998), hep-th/9809125; hep-th/9809127.
\item \label{bib:planck} M.\ Planck, {\it Theory of Heat Radiation}
(Dover, New York, 1959).
\item \label{bib:messiah} A.\ Messiah, {\it Quantum Mechanics}
Vol.\ 1 (North Holland, New York, 1961) p.\ 232.
\item \label{bib:prd34} E.\ R.\ Floyd, {\it Phys.\ Rev.}\ {\bf D
34}, 3246 (1986).  
\item \label{bib:afb20} E.\ R.\ Floyd, {\it An.\ Fond.\ Louis de
Broglie} {\bf 20}, 263 (1995). 
\item \label{bib:carroll} R.\ Carroll, {\it Can.\ J.\ Phys.}\ {\bf
77}, 319 (1999), quant-ph/9903081
\item \label{bib:bohr} L.\ Rosenfeld, ed.\ {\it Niels Bohr,
Collected Works} Vol.\ 3 (North Holland, New York, 1976)
\item \label{bib:thooft} G. 't~Hooft, {\it Class.\ Quant.\ Grav.}\
{\bf 16}, 3263 (1999), gr-qc/9903084
\item \label{bib:mpc} M.\ Matone, private communication.
\item  \label{bib:pr85} D.\ Bohm, {\it Phys.\ Rev.}\ {\bf 85}, 166
(1952).
\item  \label{bib:pr144} D.\ Bohm and B.\ J.\ Hiley, {\it Phys.\
Rep.}\ {\bf 144}, 323 (1987).
\item \label{bib:prd26} E.\ R.\ Floyd, {\it Phys.\ Rev.}\ {\bf D
26}, 1339 (1982). 
\item \label{bib:prd29} E.\ R.\ Floyd, {\it Phys.\ Rev.}\ {\bf D
29}, 1842 (1984).
\item \label{bib:pe7} E.\ R.\ Floyd, {\it Phys.\ Essays}\ {\bf 7},
135 (1994).
\item \label{bib:fpl9} E.\ R.\ Floyd, {\it Found.\ Phys.\ Lett.}\
{\bf 9}, 489 (1996), quant-ph/9707051.
\item \label{bib:d524} H.\ B.\ Dwight, {\it Table of Integrals and
other Mathematical Data} 4th ed.\ (McMillan, New York, 1961) \P
858.524.
\item \label{bib:d534} {\it ibid}., \P 858.534.
\item \label{bib:pra57} E.\ Recami and G.\ Salesi, {\it Phys.\
Rev.}\ {\bf A 57}, 98 (1998). 
\item \label{bib:qp9902019} S.\ Esposito, {\it Found.\ Phys.\
Lett.}\ {\bf 12}, 165 (1999), quant-ph/9902019.
\item \label{bib:maslov} V.\ P.\ Maslov and M.\ V.\ Fedoriuk, {\it
Semi-Classical Approximation in Quantum Mechanics} (D.\ Reidel,
Boston, 1981).
\item \label{bib:ap166} G.\ Barton, {\it Ann.\ Phys.\ (NY)} {\bf
166}, 322 (1986).
\item \label{bib:qp9708007} E.\ R.\ Floyd, quant-ph/9708007.
\item \label{bib:ajp60} J.\ Ford and G.\ Mantica, {\it Am.\ J.\
Phys.}\ {\bf 60}, 1086 (1992). 
\item \label{bib:ijmpa14} E.\ R.\ Floyd, {\it Int.\ J.\ Mod.\
Phys.}\ {\bf A 14}, 1111 (1999), quant-ph/9708026.
\item \label{bib:park} D.\ Park, {\it Classical Dynamics and Its
Quantum Analogues}, 2nd ed.\ (Springer-Verlag, New York, 1990) p.\
142.
\item \label{bib:bohm-qm} D.\ Bohm, {\it Quantum Mechanics}
(Prentice-Hall, Englewood Cliffs, 1951) p.\ 100.
\item \label{bib:bohm-qm-2} {\it ibid.}, p.\ 28.
\end{enumerate} 

\end{document}